\documentclass[%
 aip,
 amsmath,amssymb,
 reprint,%
]{revtex4-1}

\usepackage{graphicx}
\usepackage{dcolumn}
\usepackage{bm}

\usepackage[utf8]{inputenc}
\usepackage[T1]{fontenc}
\usepackage{mathptmx}
\usepackage{etoolbox}

\usepackage{array}
\usepackage{subfigure}

\usepackage{xcolor}
\usepackage[]{mathastext}
\usepackage{subfigure}

\makeatletter
\def\@email#1#2{%
 \endgroup
 \patchcmd{\titleblock@produce}
  {\frontmatter@RRAPformat}
  {\frontmatter@RRAPformat{\produce@RRAP{*#1\href{mailto:#2}{#2}}}\frontmatter@RRAPformat}
  {}{}
}%
\makeatother

\begin{document}

\title{Domain wall and Magnetic Tunnel Junction Hybrid for on-chip Learning in UNet architecture}

\author{Venkatesh Vadde$^\%$}
\affiliation{Department of Electrical Engineering, Indian Institute of Technology Bombay, Powai, Mumbai-400076, India}%
\author{Bhaskaran Muralidharan$^{\$}$}%

\affiliation{Department of Electrical Engineering, Indian Institute of Technology Bombay, Powai, Mumbai-400076, India}%

\author{Abhishek Sharma$^\#$}
\affiliation{Department of Electrical Engineering, Indian Institute of Technology Ropar, Rupnagar, Punjab-140001, India}%

\email{$^\%$vaddevenkatesh19@gmail.com, $^\$$bm@ee.iitb.ac.in, $^\#$abhishek@iitrpr.ac.in}

\begin{abstract}
We present spintronic devices based hardware implementation of UNet for segmentation tasks. Our approach involves designing hardware for convolution, deconvolution, rectified activation function (ReLU), and max pooling layers of the UNet architecture. 
We designed the convolution and deconvolution layers of the network using the synaptic behavior of the domain wall MTJ. We also construct the ReLU and max pooling functions of the network utilizing the spin hall driven orthogonal current injected MTJ. 
To incorporate the diverse physics of spin-transport, magnetization dynamics, and CMOS elements in our UNet design, we employ a hybrid simulation setup that couples micromagnetic simulation, non-equilibrium Green's function, SPICE simulation along with network implementation.
We evaluate our UNet design on the CamVid dataset and achieve segmentation accuracies of 83.71$\%$ on test data, on par with the software implementation with 821mJ of energy consumption for on-chip training over 150 epochs. We further demonstrate nearly one order $(10\times)$ improvement in the energy requirement of the network using unstable ferromagnet ($\Delta$=4.58) over the stable ferromagnet ($\Delta$=45) based ReLU and max pooling functions while maintaining the similar accuracy. The hybrid architecture comprising domain wall MTJ and unstable FM-based MTJ leads to an on-chip energy consumption of 85.79mJ during training, with a testing energy cost of 1.55 $\mu J$.

\end{abstract}
\maketitle
\section{Introduction}
Semantic image segmentation is a pixel-level classification of an image and involves clustering parts of the image that belong to the same class\cite{athanasiadis2007semantic, liu2019recent, thoma2016survey}. This deep learning task is integral to computer vision and pattern recognition and has substantial use in fields such as medical imaging\cite{asgari2021deep}, self-driving cars\cite{cui2021deep}, and satellite imagery analysis\cite{huang2002assessment}.
While convolutional neural networks (CNNs) like LeNet, VGGNet, and GoogleNet are commonly used for classification tasks, where the output is a single class label, semantic segmentation requires localization information, i.e., a class label for each pixel. Consequently, image segmentation, with its pixel-wise classification, is computationally more demanding than object classification. 
The architectures for image segmentation, like UNet, play a crucial role in diffusion models used for image generation, such as OpenAI's DALL-E. This emphasizes the necessity of developing hardware implementations for these networks.

Implementing these complex deep neural network algorithms on traditional hardware which are based on von-Neumann architecture is resource-intensive in terms of energy consumption, area, and time. This is primarily due to the separation of memory and processing units.  So, there is a need for specialized hardware designs that utilize in-memory computing paradigm\cite{grollier2020neuromorphic}, offering optimization tailored for the efficient implementation of deep neural networks.

Several studies have investigated the specialized hardware implementation of segmentation tasks\cite{menon2021high,liu2018optimizing}. These works are based on optimizing the segmentation for FGPA implementation\cite{liu2018optimizing} and deploying a pipelined VLSI architecture\cite{menon2021high}. These works are based on CMOS devices and consume high power and area. Spintronic devices on the other hand consume lower power, area\cite{grollier2020neuromorphic} and are compatible with CMOS technology\cite{chung20164gbit}. Spintronic devices also have the advantage of having a diverse range of properties such as non-volatility, oscillatory, plasticity, high endurance, linear response, and stochastic behavior \cite{grollier2020neuromorphic,vadde2023orthogonal, sharma2017resonant, hirohata2020review, camsari2017implementing}.
These properties give a wide range of tools to design specialized hardware for deep neural network implementation. 
While spintronic realizations of multilayer perceptrons\cite{sengupta2016proposal,grollier2020neuromorphic}, convolutional neural networks\cite{desai2022chip}, spiking neural networks\cite{das2023bilayer} and reservoir computing\cite{markovic2019reservoir} have been demonstrated, the implementation of UNet remains elusive.
In this work, we propose a spintronic implementation of convolution, deconvolution, ReLU, and max-pooling layers that are essential for UNet. A hybrid of domain-wall MTJ and SHE-MTJs are employed for realizing these layers. We utilize a hybrid simulation method that couples micromagnetic simulation, Keldysh non-equilibrium Green’s (NEGF) function, 
and SPICE simulation with network implementation to capture the diverse physics of spintronic and CMOS devices in our designs.

\begin{figure*}[!ht]
	\centering
     \includegraphics[width=0.99\linewidth]{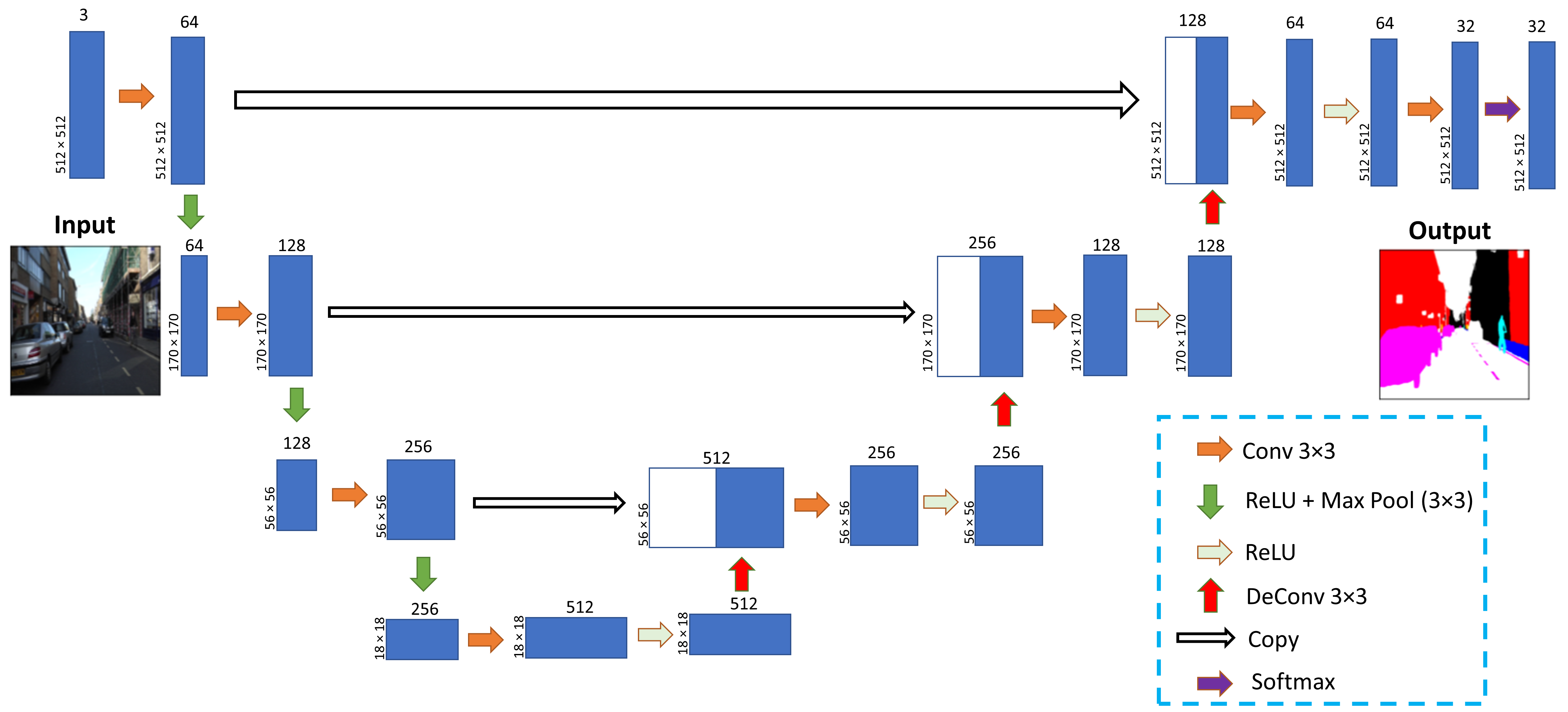}
	\quad
	\caption{The UNet structure is illustrated, where the feature map is represented by blue boxes with the number of channels indicated at the top and the size displayed on the left edge. White boxes signify copied features from previous stages, and arrows indicate various operations. An example input image and its corresponding output are also depicted.}
	\label{fig:unet_sche}
\end{figure*}
The rest of the paper is organized as follows, in section \ref{sect:unet_design}, we describe the UNet architecture used for image segmentation and explain how convolution and deconvolution can be realized using cross-bar arrays and the characteristics of ReLU and max pooling layers.
Section \ref{sect:simulation_method} delves into the simulation method, outlining the coupling of micromagnetic simulation, NEGF, and circuit simulation with network implementation to execute image segmentation.
In section \ref{sect:domain_wall_synapse}, we describe the domain-wall MTJ and discuss the synaptic behavior of the domain-wall device.
In section \ref{sect:relu_max}, we present the orthogonal current injected MTJ device and circuit designs for ReLU and max pooling functions.
In section \ref{sect:segmen_results}, we show the results of the image segmentation using the CamVid dataset and compare the on-chip energy consumption of the proposed network for different thermal stability factors.
In section \ref{sect:discussion}, we discuss the possibility of physical spintronic realization of complex networks with large numbers of parameters.
We conclude in section \ref{sect:conclusion}.

\section{Architecture for Segmentation}
\label{sect:unet_design}

There are multiple architectures developed for image segmentation like UNet, SegNet, etc. Among these, UNet has been widely adopted in segmentation tasks.
The UNet architecture was initially proposed by Olaf Ronneberger et al\cite{ronneberger2015u} for medical image segmentation. This architecture consists of two main components: the contracting path (also known as the encoder) and the expanding path (also known as the decoder), connected by a copy path (also referred to as skip connection). The contracting path reduces the feature map while extracting image features, and the expanding path utilizes these features to localize objects and reconstruct the segmentation mask\cite{ronneberger2015u}. As the feature map undergoes a reduction in the contracting path, some information is lost, to address this, the copy connection (skip connection) is employed to reintroduce the lost information to the expanding path.
Figure \ref{fig:unet_sche} shows the schematic of a UNet structure, here the contracting path contains convolution, ReLU, and Max-pooling layers while the expanding path contains deconvolution, convolution, and ReLU layers terminated by a softmax function. In this network, we employ 4.65 million domain-wall synapses, 21.45 million ReLU circuit instances, and 2.33 million ReLU-Max pooling circuit instances to tackle the highly complex task of semantic image segmentation.

Implementing image segmentation through UNet on hardware necessitates the design of circuits dedicated to convolution, deconvolution, ReLU activation functions, and max-pooling layers. In the following sections, we describe the networks designed for these layers.

\subsection{Convolution}
\begin{figure*}[!ht]
	\centering
     \includegraphics[width=0.9\linewidth]{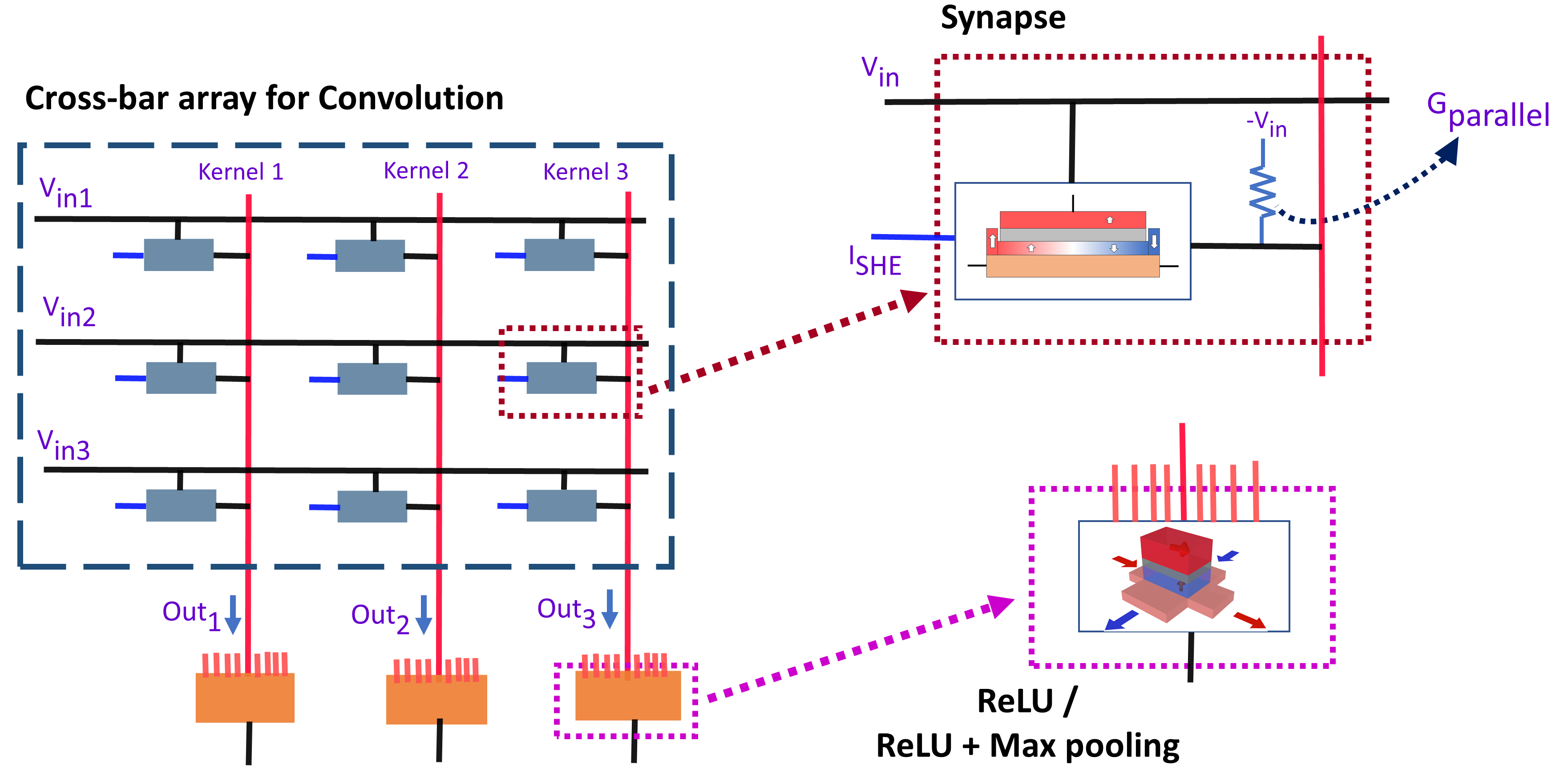}
	\quad
	\caption{The convolution operation using DW-based cross-bar array. The vertical lines symbolize the convolution kernels, and the input is applied to the horizontal lines. The DW device along with parallel conductance is used to store the kernel values. The resulting current output from the vertical lines (kernel output) is connected to the ReLU/ReLU+Max pooling devices.}
	\label{fig:crossbar}
\end{figure*}

The convolution operation entails matrix-vector multiplication, where the input is multiplied with a kernel. This matrix-vector multiplication operation is fundamental to artificial neural networks where the input/feature map is multiplied with a weight matrix. In convolution, the kernel can be thought of as a weight matrix. Performing this vector multiplication requires a lot of memory fetches when using traditional hardware based on von-Neumann architecture. So, crossbar arrays\cite{xia2019memristive, jung2022crossbar, desai2022chip} have become very popular for matrix-vector multiplication, an example of a crossbar array is shown in Fig. \ref{fig:crossbar}. In crossbar arrays, the weight matrix/kernels are stored in non-volatile memory elements (synapses), where analog memory and computing units are intricately interwoven, leading to faster and more energy-efficient matrix multiplication. In Fig. \ref{fig:crossbar}, the inputs are applied to horizontal lines, while the kernel weights are stored as conductances of synapses in the vertical lines, and the output of the vector multiplication(weighted sum of inputs) is given by the current value in the vertical lines.

To implement such a crossbar array, a non-volatile synaptic device is necessary. Therefore, we employ a domain-wall-based magnetic tunnel junction (DW-MTJ) device to store the kernel weight. The neural network can have both positive and negative weights, but the conductance values of the DW-MTJ are positive only. To address this we add a conductance in parallel to the DW-MTJ as shown in Fig. \ref{fig:crossbar}. So the weight can be represented as 
\begin{equation}
    W_{i,j} = G_{DW MTJ} - G_{parallel}
\end{equation}

\begin{equation}
    G_{parallel} = \frac{G_{AP} + G_{P}}{2}
\end{equation}
Here, $W_{i,j}$ is the weight connecting $i^{th}$ input with $j^{th}$ kernel, $G_{DW MTJ}$ is the conductance of the DW-MTJ. $G_{AP}$ and $G_{P}$ are the anti-parallel and parallel conductances of the DW-MTJ.
Further details about the DW-MTJ device are elaborated in Section \ref{sect:domain_wall_synapse}.

\subsection{Deconvolution}
Deconvolution, also referred to as transposed convolution or fractionally-strided convolution, operates in the reverse direction of convolution. It extrapolates new information from the feature map and can be thought of as a one-to-many connection\cite{liu2018optimizing}. Deconvolution serves as a technique for upsampling images, resulting in an output size larger than the input size. This operation has significant application in generative adversarial networks and fully convolutional networks\cite{li2020red}.

\begin{figure}[!ht]
	\centering
     \includegraphics[width=0.99\linewidth]{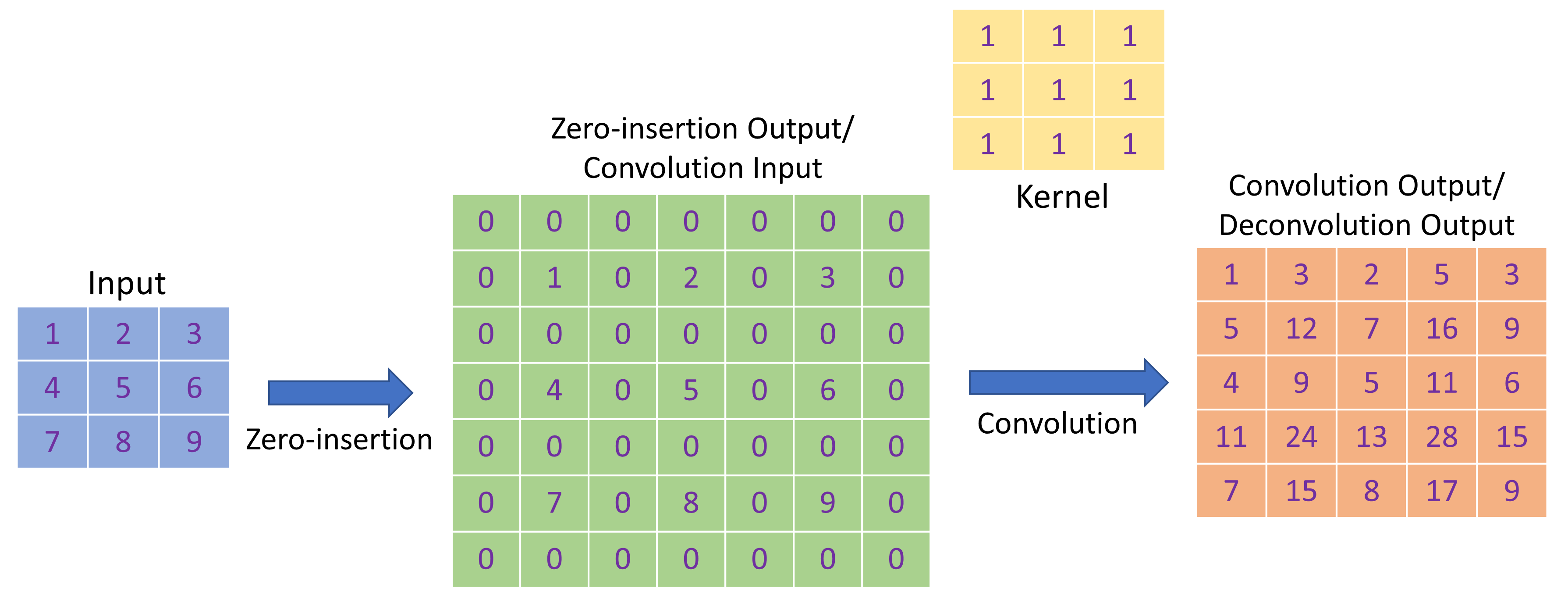}
	\quad
	\caption{The deconvolution operation as a combination of zero-insertion and convolution operation.}
	\label{fig:deconv}
\end{figure}

The deconvolution operation can be achieved by introducing zeros into the input matrix and performing a convolution operation\cite{chen2018regan,li2020red}. Figure \ref{fig:deconv} illustrates the deconvolution operation as a combination of zero insertion and convolution. Zeros are inserted along each row and column, including at the edges of the input matrix, thereby expanding the input size. This up-sampled matrix is then used as input for convolution. This combination of zero insertion and convolution yields the same effect as deconvolution. While this method involves redundant operations of multiplication with zeros, it allows us to utilize the convolution operation for which we have designed a hardware implementation using cross-bar arrays in the previous section. This approach reduces the complexity of the hardware design for the segmentation tasks. Hence, in our network design, we represent deconvolution through the convolution operation with an additional step of zero insertion.

\begin{figure*}[!ht]
	\centering
     \includegraphics[width=0.999\linewidth]{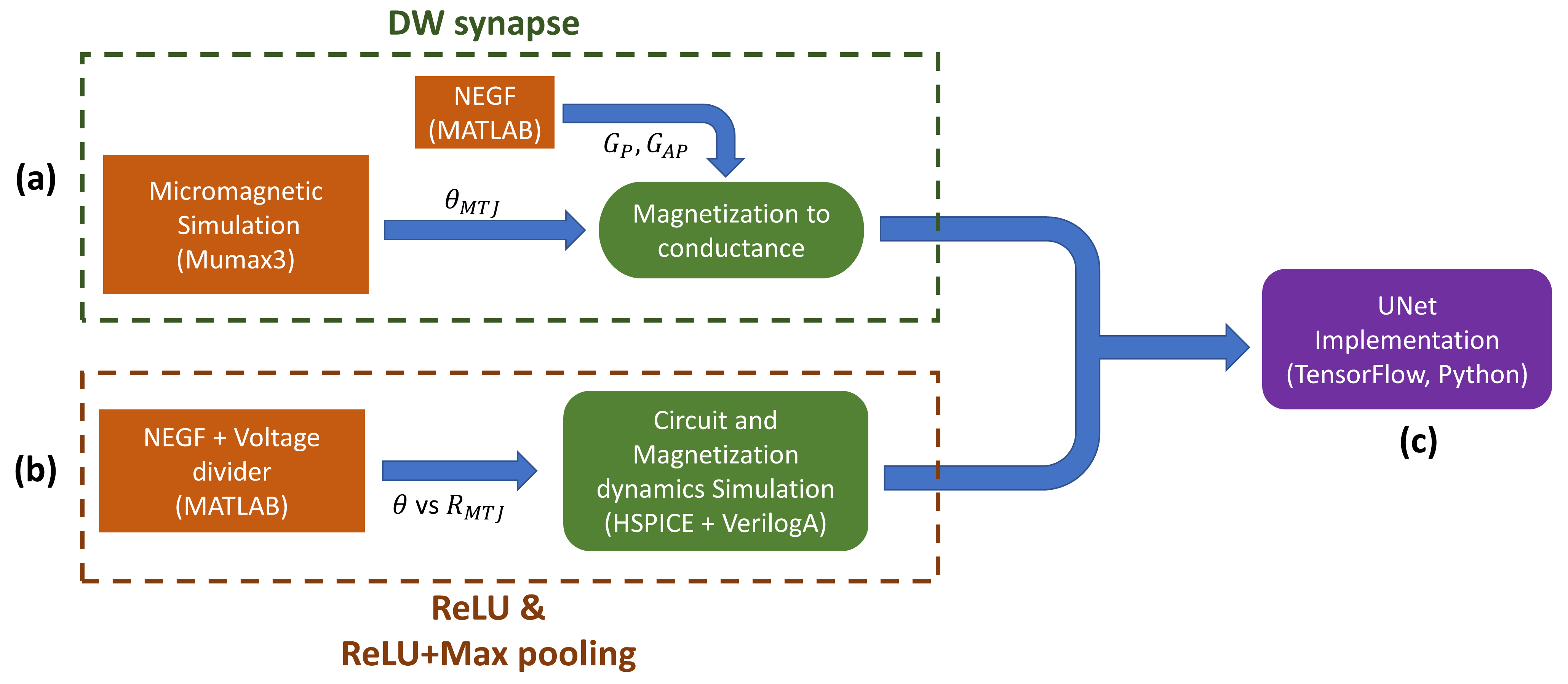}
	\quad
	\caption{Overview of the simulation setup. (a) Micromagnetic simulation of the domain wall is simulated in mumax3, and the magnetization outcomes are translated to MTJ conductance using parallel and anti-parallel conductances obtained from NEGF simulation. (b) Hybrid NEGF-CMOS simulation setup for ReLU and ReLU-max pooling circuits. The NEGF is interconnected with a voltage divider circuit in a self-consistent manner to compute MTJ resistance. This resistance is then integrated into HSPICE circuit simulation using VerilogA. The LLGS equation is interconnected with other circuit components to compute ReLU and ReLU-max pooling functions. (c) The characteristics of the DW synapse, ReLU circuit, and ReLU-max pooling network are incorporated into the TensoFlow package to implement the UNet architecture, which is utilized for semantic image segmentation. }
	\label{fig:sim_method}
\end{figure*}

\subsection{ReLU and Max-pooling}
Activation functions play a crucial role in neural networks, introducing non-linearity that enables the network to learn intricate structures and distinguish between outputs\cite{goodfellow2016deep}. The rectified linear activation function (ReLU)\cite{vadde2014she_mmm} has emerged as a default choice for various networks, as it has been shown to improve learning in neural networks\cite{glorot2011deep, ide2017improvement, nair2010rectified}.
In convolutional neural networks (CNNs), UNet, and fully connected convolutional networks, a pooling layer is commonly incorporated to reduce the size and parameters while extracting features. Among various pooling methods, max pooling is popular, max pooling also has the ability to suppress noise by discarding noisy activations\cite{goodfellow2016deep}.

To implement the ReLU function, we employ an orthogonal current-injected MTJ design. Subsequently, we utilize this ReLU circuit to construct a $3\times3$ max pooling network that simultaneously performs both ReLU and max-pooling functions.  We discuss these implementations in Section \ref{sect:relu_max}.

\section{Simulation method}
\label{sect:simulation_method}

\begin{figure*}[!ht]
	\centering
 
	\subfigure[]{\includegraphics[width=0.32\linewidth]{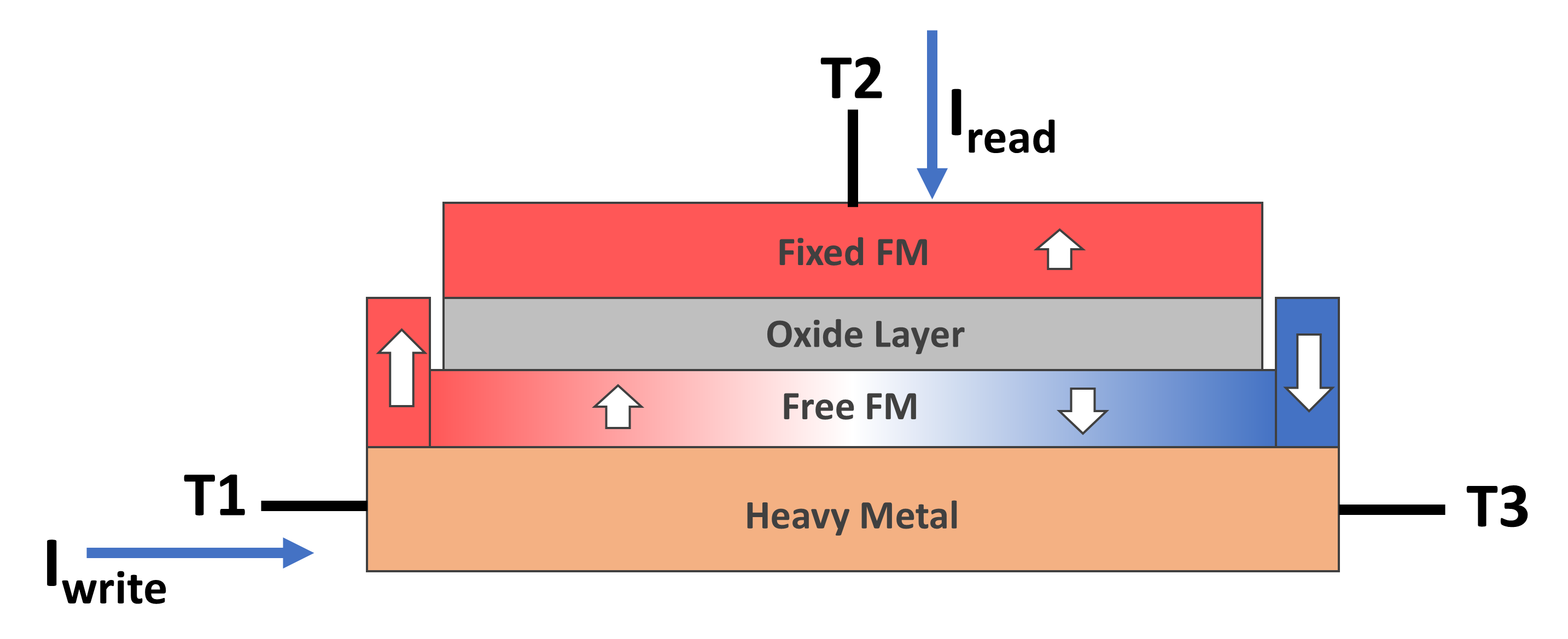}\label{fig:DW_schema}}
	\subfigure[]{\includegraphics[width=0.33\linewidth]{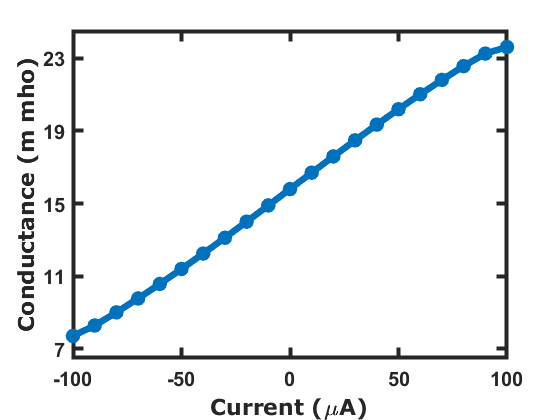}\label{fig:DW_cond}}
 \subfigure[]{\includegraphics[width=0.33\linewidth]{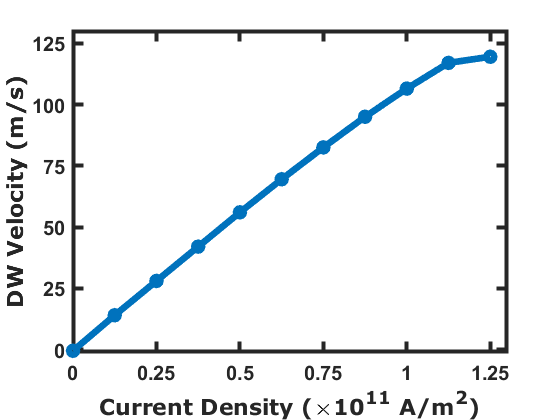}\label{fig:DW_velo}}
	\caption{(a) Schematic of the domain-wall based synapse.  $I_{write}$ denotes the write current passing through terminals T1 and T3, while $I_{read}$ represents the read current flowing through terminals T2 and T3. (b) The conductance of the DW-MTJ device with respect to input current pulse ($I_{write}$). (b) The velocity of the domain wall with varying input current density.}
\end{figure*}

Figure \ref{fig:sim_method} presents an overview of the simulation method, encompassing micromagnetic simulation, NEGF formalism, magnetization dynamics, circuit simulation, and UNet implementation. The simulation can be divided into three components: domain-wall synapse simulation, ReLU-max pooling design, and UNet implementation.

The implementation of the domain-wall synapse involves micromagnetic simulation, which gives the response in magnetization of the free ferromagnetic layer due to applied current. To perform these micromagnetic simulations, the mumax3 software\cite{vansteenkiste2014design, mulkers2017effects} was employed. These magnetization results are used to obtain the conductance of DW-MTJ devices using the following equation \cite{sengupta2016proposal, bhowmik2019chip}.

\begin{equation}
    G_{DW-MTJ} = G_P cos^2(\frac{\theta}{2}) + G_{AP} sin^2(\frac{\theta}{2}) 
\end{equation}
Here, $\theta$ represents the angle between free-FM and fixed-FM magnetizations, $G_P$ and $G_{AP}$ are the parallel and anti-parallel conductances of the MTJ. NEGF simulation is utilized to compute the $G_P$ and $G_{AP}$ conductance values.

In the simulation of the ReLU-Max pooling network, NEGF simulation is self-consistently coupled with the voltage divider(formed by MTJ and a fixed resistor). Here the MTJ angle is varied to find the resistance of the MTJ by iteratively calculating the voltage across the MTJ. The results from the NEGF simulation are incorporated into the HSPICE circuit simulator through VerilogA, where the Verilog-A component provides the MTJ resistance based on the MTJ angle given by HSPICE. The HSPICE also performs magnetization dynamics simulation to find the MTJ angle along with the CMOS device simulations based on the 16nm predictive technology model\cite{Predicti66:online}. The results of the domain-wall synapse, ReLU, and max pooling are incorporated into the Python programming, where the UNet architecture shown in Fig.\ref{fig:unet_sche} is implemented using the TensorFlow package. 
For implementing the UNet architecture in python, we utilize the conductance relationship of the domain-wall MTJ derived from mumax and NEGF, for the ReLU and ReLU-max pooling circuits, we use the empirical relationship between input current and output voltage along with the performance in the presence of thermal noise obtained from HSPICE and NEGF.

\subsection{Quantum transport: NEGF}
We use the Keldysh NEGF technique \cite{camsari2020non, datta2011voltage, sharma2017resonant} to simulate the transport through MTJ that has MgO sandwiched between free and fixed CoFeB FM layers. The NEGF formalism is given by

\begin{gather}
    G(E) = [EI-H-\Sigma]^{-1},\\
    [H]=[H_0]+[U]\\
    \Sigma=\Sigma_T + \Sigma_B\\
    G^n= \int dE [G(E)] [\Sigma^{in}(E)][G(E)]^\dagger
\end{gather}

Here $G(E)$ is the Green's function matrix, $[I]$ is the identity matrix, $E$ is the energy variable, 
$[H]$ is the device Hamiltonian, $[H_0]$ is the device tight-binding matrix, $[U]$ is the Coulomb charging matrix, 
$\Sigma$ is the self-energy matrix and $\Sigma_{T,B}$ are the self-energy matrices for the top (fixed) and bottom (free) FM layers respectively. 
$G^n$ is the electron correlation matrix and $\Sigma^{in}$ is the in-scattering function. 

The quantum transport part leads to the calculation of the current operator ($I_{op}$) that represents the charge current between two lattice points i and i+1 is given by

\begin{equation}
    I_{op} = \frac{i}{\hbar}(H_{i,i+1}G^n_{i+1,i} - H_{i+1,i}G^n_{i,i+1})
\end{equation}

\begin{equation}
    I= q \int Real [Trace(\hat{I}_{op})] dE
\end{equation}

The current operator $I_{op}$ is $2 \times 2$ matrix in the spin space of the lattice point. Here $I$ is the charge current through the MTJ device and $q$ is the quantum of electronic charge.

\subsection{Magnetization dynamics}
The Landau-Lifshitz-Gilbert-Slonczewski (LLGS) equation \cite{slonczewski1996current, panagopoulos2013physics} is used to describe the magnetization dynamics of the free-FM. The LLGS equation is given by
\begin{multline}
  (\frac{1+\alpha^2}{\gamma H_k})\frac{d \hat{m}}{dt} = -\hat{m} \times \Vec{h}_{eff} - \alpha \hat{m} \times \hat{m} \times \Vec{h}_{eff} \\ - \hat{m} \times \hat{m} \times \Vec{i}_s + \alpha \hat{m} \times \Vec{i}_s,
\end{multline}
where $\hat{m}$ is the unit vector along the direction of magnetization of the free magnet, $\gamma$ is the gyromagnetic ratio, $\alpha$ is the Gilbert damping parameter, $\Vec{h}_{eff} = \frac{\Vec{H}_{eff}}{H_k}$ is the reduced effective field and $\Vec{i}_s= \frac{\hbar\Vec{I}_{s}}{2qM_sVH_k}$ is the normalized spin current. The term $\vec{H}_{eff}$ includes the contribution of the anisotropy field ($H_k$) and the thermal noise ($H_{th}$). The thermal noise\cite{sun2004spin} is given by $ \langle H_{th}^2 \rangle = \frac{2\alpha k_B T}{\gamma M_s V}$ and $\langle \rangle$ represents the ensemble average. 

\begin{figure*}[!ht]
	\centering
     \includegraphics[width=0.99\linewidth]{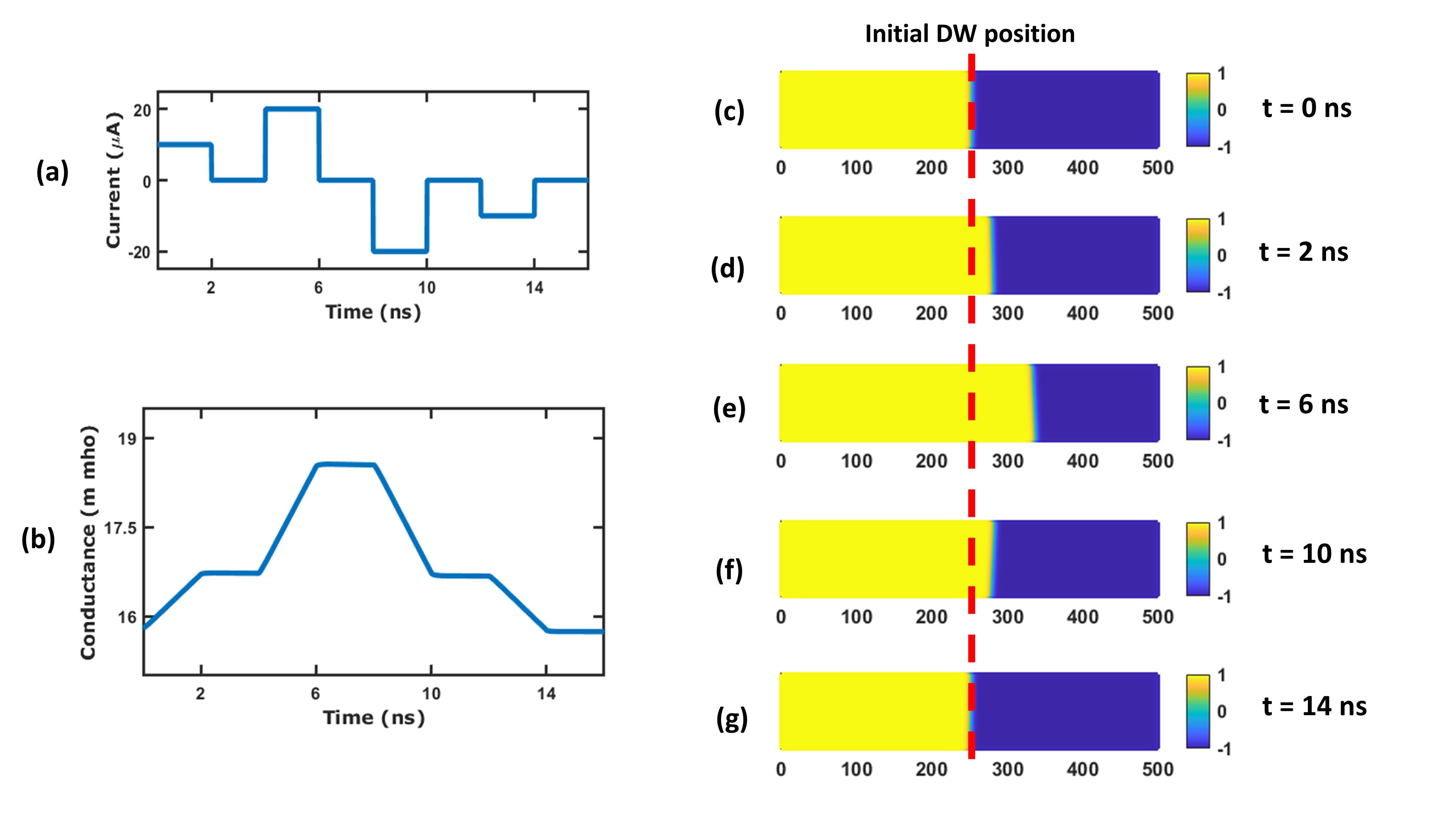}
	\quad
	\caption{Response of the domain-wall to a write current pulse. (a) Train of write current pulses. (b) Conductance of the DW-MTJ corresponding to the current pulse train. (c) $m_z$ magnetization of the free-FM of the DW-MTJ at t=0 ns. (d) - (g) Snapshots of the $m_z$ magnetization at various time points, illustrating the response to the current pulse train.}
	\label{fig:DW_pulse}
\end{figure*}

\subsection{SHE layer}
The charge-to-spin conversion via the spin hall effect(SHE) in heavy metals is used to effectively manipulate the free-FM magnetization. The charge-to-spin conversion of the SHE layer and the polarization of the generated current is given by \cite{song2021spin,liu2012current,takahashi2008spin} 
\begin{equation}
    \theta _{SH} = \frac{J_s}{J_c}
\end{equation}
\begin{equation}
    I_s = \theta_{SH} \frac{L}{t} I_c 
    \label{eq:Ic_Is}
\end{equation}
\begin{equation}
    \hat{I_s} = \hat{I_c} \times \sigma
    \label{eq:she_dir}
\end{equation}

Here, $J_s$ is the spin current density and $J_c$ is the charge current density.
$I_s$ is the spin current generated, $\theta_{SH}$ is the spin Hall angle of the heavy metal, L, t are the length and thickness of the heavy metal, and $I_c$ is the charge current injected. $\hat{I_s}$ is the direction of generated spin current flow, $\hat{I_c}$ is the direction of input charge current, and $\sigma$ is the polarization of the generated spin current. From Eq. \ref{eq:she_dir}, injection of charge current to heavy metal in $\hat{x}-$ direction results in y-polarized spin current injection to the free-FM (z-direction) on top of the HM layer.

The resistance (R) of the heavy metal is given by
\begin{equation}
    R=\rho \frac{L}{W t}
\end{equation}
Here, $\rho$ and W are the resistivity and width of the heavy metal respectively.

\section{Domain wall Synapse}
\label{sect:domain_wall_synapse}

The domain wall synapse is a 3-terminal device as shown in Fig. \ref{fig:DW_schema}. In this 3-terminal configuration, the read and write paths are distinct, preventing accidental modification of synapse information during reading\cite{sengupta2016proposal, bhowmik2019chip}. The write path in the DW device, illustrated in Fig. \ref{fig:DW_schema}, is between terminals T1 and T3, while the read path is between terminals T2 and T3.
The free-FM layer of the DW-MTJ has two oppositely polarized magnetic regions separated by a domain wall. This domain wall can be moved by spin orbit torque (SOT) exerted by the heavy metal. Thus the charge current flowing through the heavy metal injects spin current into the free-FM layer and moves the domain wall. The two pinned layers on either side of the free FM layer help prevent the domain wall from getting destroyed when a high current is applied. The movement of the domain wall causes one magnetic region to shrink while the other expands this changes the average magnetization of the free-FM layer. This change in magnetization translates to a variation in the conductance of the device, due to the tunnel magneto-resistance effect of the MTJ. 

\begin{figure*}[!ht]
	\centering
 
	\subfigure[]{\includegraphics[width=0.28\linewidth]{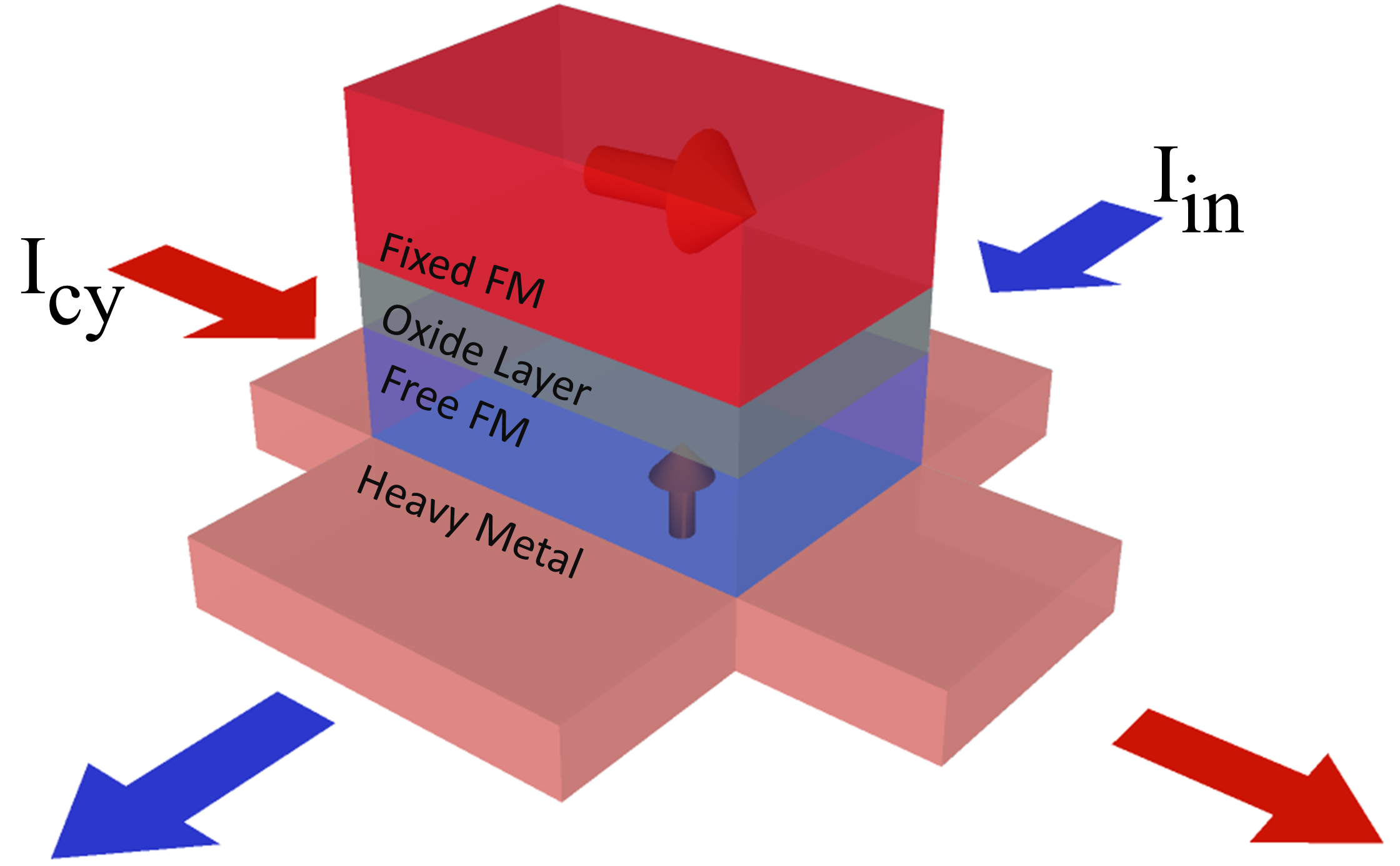}\label{fig:relu_3d}}
	\subfigure[]{\includegraphics[width=0.33\linewidth]{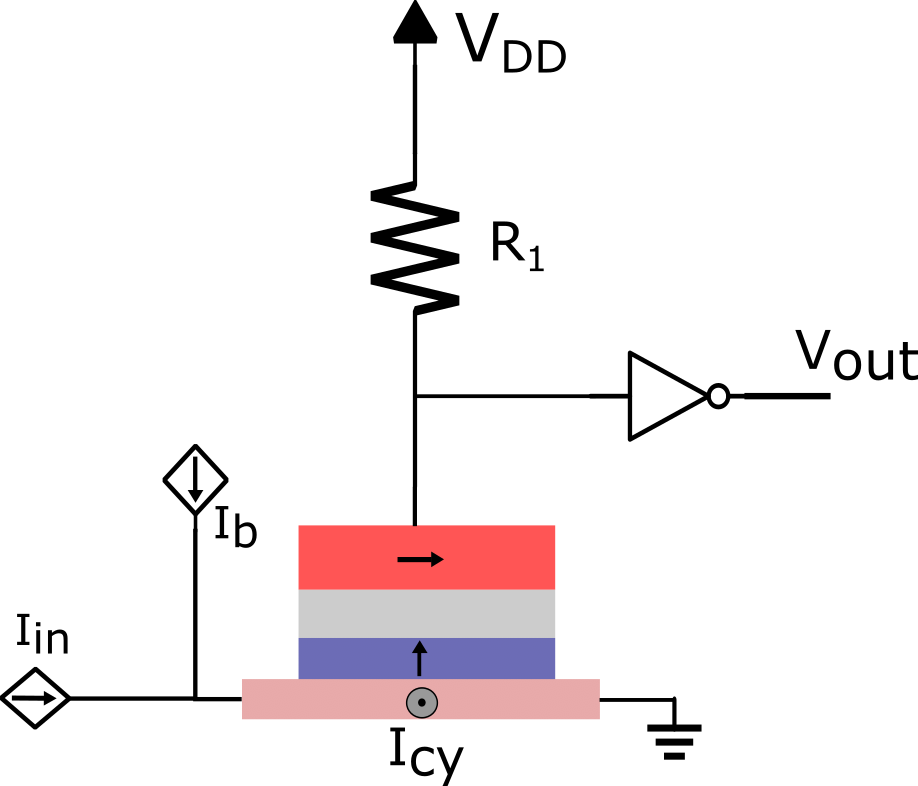}\label{fig:relu_cir_schema}}
 \subfigure[]{\includegraphics[width=0.33\linewidth]{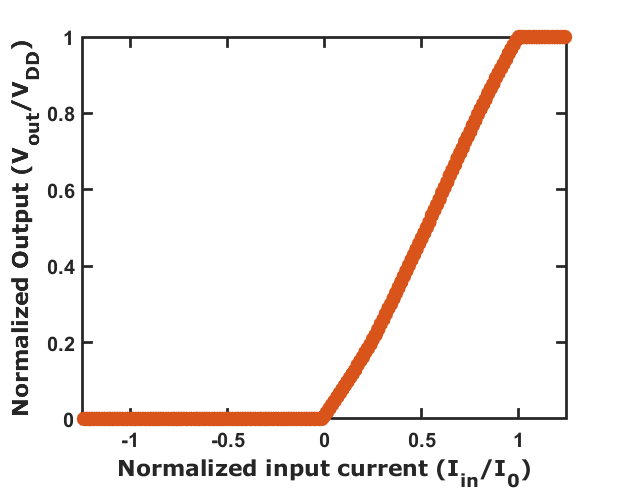}\label{fig:relu_vout}}
	\caption{(a) Schematic of the orthogonal current injected SHE-MTJ device for continuous resistance change. (b) Circuit design for ReLU function emulation. (c) The output of the ReLU circuit with $I_0 = 14.5 \mu A$, $V_{DD}=0.5V$  and $\Delta=4.58$. }
\end{figure*}

\subsection{Device parameters}
The spin orbit coupling at the heavy metal-free FM interface leads to Dzyaloshinskii-Moriya exchange interaction (DMI) which stabilizes the Neel domain wall\cite{emori2013current, martinez2014current, shahbazi2018magnetic, sengupta2016proposal, bhowmik2019chip}. For our synaptic device, we consider a PMA CoFeB ferromagnet with dimensions $500 \times 100 \times 1$ $nm$, saturation magnetization($M_s$) of 0.7 $MA/m$, PMA constant($K_u$) of 0.8 $MJ/m^3$, exchange-correlation constant($A_{ex}$) of 10 $pJ/m$, damping constant($\alpha$) of 0.3 and DMI constant($D$) of 1.2 $mJ/m^2$. We consider the highly efficient $Au_{0.25}Pt_{0.75}$ heavy metal \cite{zhu2018highly, vadde2023power}, with spin hall angle ($\theta_{SHE}$) of 0.3, resistivity ($\rho$) of 83 $\mu \Omega cm$ and a thickness of 4nm, resulting in a resistance of 1037.5 $\Omega$.  $Au_{0.25}Pt_{0.75}$ is taken as a heavy metal since it has a low spin Hall power factor\cite{vadde2023power} so it is more power efficient compared to other heavy metals.

\subsection{Results}
We show in Fig. \ref{fig:DW_cond} the linear conductance relation of the DW-MTJ with the input write current pulse. The domain wall's initial position is at the center, and a current pulse of 2ns duration is applied. We observed that a 100 $\mu A$ current pulse is needed to move the domain wall to the right edge starting from the center, corresponding to the parallel alignment with the fixed FM layer, and -100 $\mu A$ is needed to move the domain wall to the left edge, corresponding to the anti-parallel alignment with the fixed FM layer. 
The velocity of the domain wall due to applied current is shown in Fig. \ref{fig:DW_velo}, which shows a linear relation for the considered parameters.

During the training of the neural network, the weights increase and decrease, so corresponding to this requirement we show in Fig. \ref{fig:DW_pulse} the response of the DW-MTJ to an input write current pulse train. Figure \ref{fig:DW_pulse}(a) shows the current pulse train, and the corresponding conductance of the DW-MTJ is shown in \ref{fig:DW_pulse}(b) over time. Figure. \ref{fig:DW_pulse}(c) - (g) show the snapshots of the free-FM magnetization showing the movement of the domain wall. We observe that the domain wall reaches its initial position over time as the net current applied is zero. We also noted the tilting of the domain wall in the presence of current, this can be explained through 1D domain wall theory\cite{martinez2014current}.

\section{ReLU and Max pooling}
\label{sect:relu_max}

\begin{figure*}[!ht]
	\centering
 
	\subfigure[]{\includegraphics[width=0.32\linewidth]{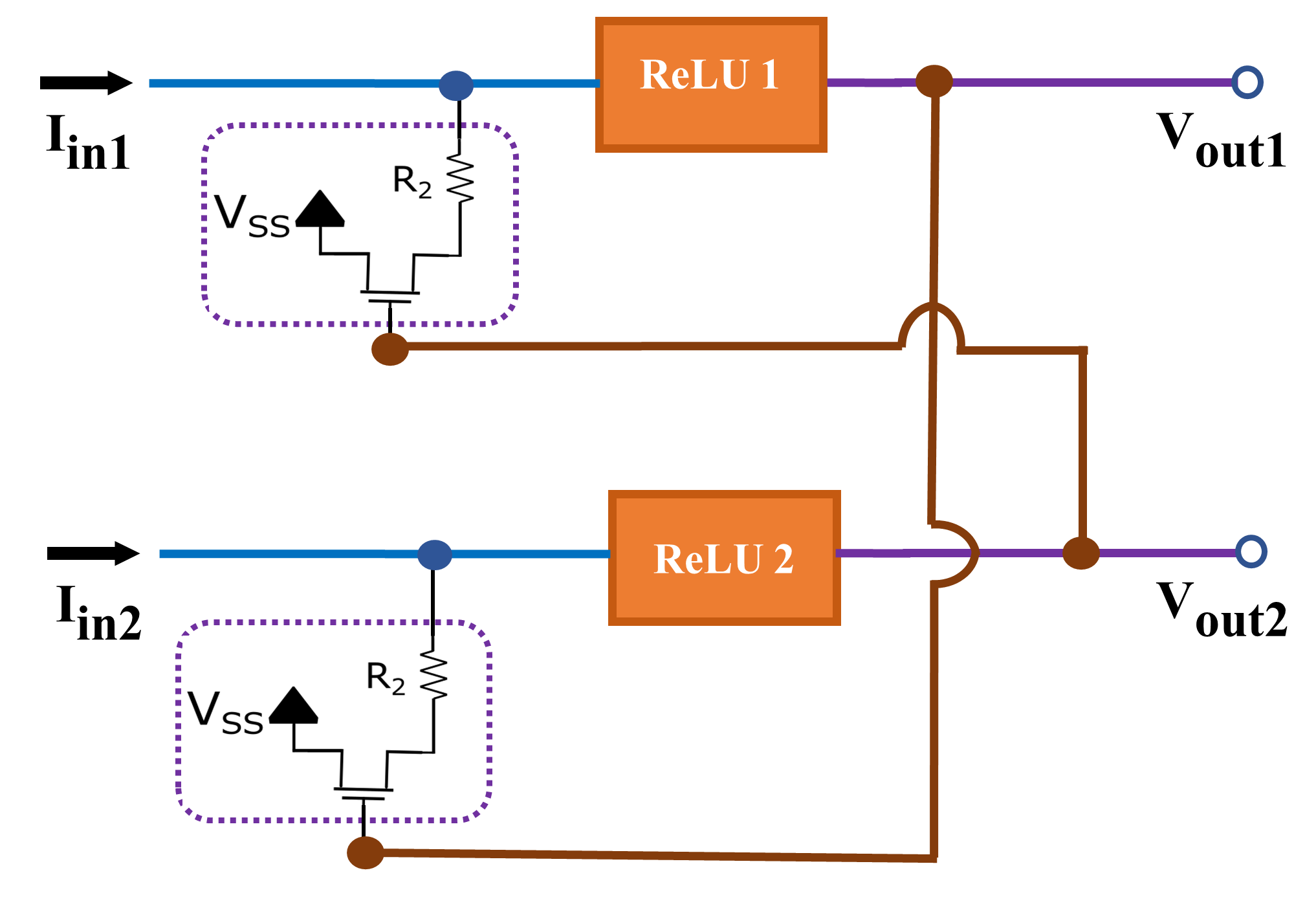}\label{fig:maxpool_schem}}
	\subfigure[]{\includegraphics[width=0.33\linewidth]{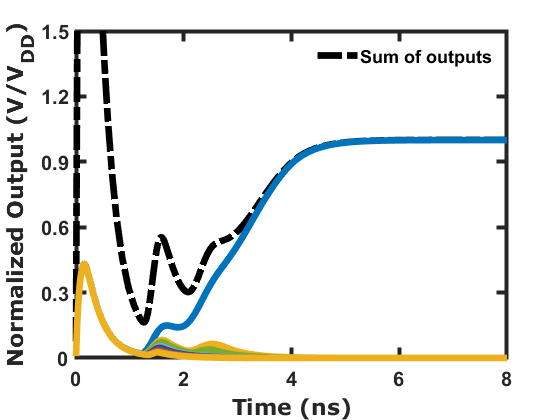}\label{fig:maxpool_tran}}
 \subfigure[]{\includegraphics[width=0.33\linewidth]{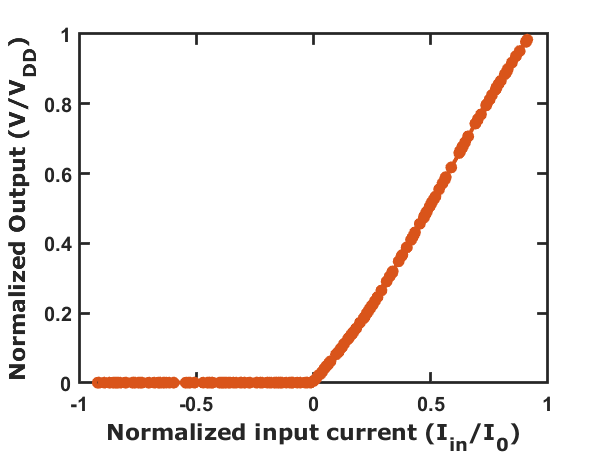}\label{fig:maxpool_out}}
	\caption{(a) Network design for ReLU-Max pooling functions. The ReLU circuits are interconnected with n-MOSFET to enable competition for max pooling functionality. (b) Transient response of the ReLU-max poling network showcasing the max pooling functionality. (c) The output of $3\times3$ ReLU-Max pooling network with $I_0 = 14.5 \mu A$, $V_{DD}=0.5V$ and $\Delta=4.58$.}
\end{figure*}

The magnetic tunnel junction possesses several properties, including the ability to undergo continuous/linear changes in resistance, which can be achieved by applying orthogonal spin currents to the free-FM of the MTJ \cite{vadde2023orthogonal}. This continuous change in resistance is essential since the ReLU function contains linear regions, requiring a device with linear characteristics for its emulation. Figure \ref{fig:relu_3d} illustrates the schematic of the MTJ injected with orthogonal spin currents generated by the SHE layer. This linear behavior of the MTJ can be utilized to construct a circuit that emulates the ReLU function, as depicted in Fig. \ref{fig:relu_cir_schema}. The circuit incorporates a resistor $R_1$, a CMOS inverter, and the MTJ device along with a current source $I_b$ to shift the output and generate the ReLU function. Injecting orthogonal currents also enhances the circuit's stability, allowing us to lower the ferromagnet's thermal stability to 4.58. This reduction helps decrease energy consumption while only slightly impacting the circuit's error.

The max pooling function entails finding the maximum of the presented inputs. To achieve this functionality we use multiple ReLU circuits and introduce competition among them so that the circuit with the highest input becomes the winner. We enable this competition through an n-MOSFET and a resistor $R_2$ connected between each pair of ReLU circuits as shown in Fig. \ref{fig:maxpool_schem}.

\subsection{Device parameters}
For the ReLU-max pooling circuits, we utilize a PMA CoFeB ferromagnet with dimensions 14.4$\times$69.4$\times$1nm. The saturation magnetization ($M_s$) is 1150$emu/cm^3$, anisotropy field ($H_k$) is 330, 2180, 3300 Oe, the Gilbert damping is 0.01, the thermal stability factor ($\Delta$) is 4.58, 30.26, 45.81, a lower thermal stability factor was used as it reduces the power consumption and our circuit design still gives accurate results \cite{vadde2023orthogonal, vadde2014she_mmm}.
The heavy metal used is $Au_{0.25}Pt_{0.75}$ \cite{zhu2018highly, vadde2023power},  with spin hall angle ($\theta_{SHE}$) of 0.3, resistivity ($\rho$) of 83$\mu \Omega cm$ and a thickness of 4nm, resulting in input resistance of 1000$\Omega$.
The circuit parameters, $I_b$ the current bias is 9.98$\mu A$, $R_1$ resistor is 698.93k$\Omega$ and the resistor $R_2$ is 16k$\Omega$.

\subsection{Results}
Figure \ref{fig:relu_vout} shows the output of the ReLU circuit, which closely resembles the ReLU activation function for normalized inputs of less than 1, here the normalized current $I_0$ is 14.5$\mu A$. The ReLU circuit consumes an average power of 0.343$\mu W$.
We show in Fig. \ref{fig:maxpool_tran} the transient results of the $3\times 3$ ReLU-Max pooling circuit, the 9 inputs are randomly taken to show max pooling functionality. Here we observe competition among the 9 ReLU circuits that enable the max pooling functionality, where the ReLU circuit with the highest input reaches its corresponding output while pushing all other ReLU units to settle to 0V. Figure \ref{fig:maxpool_out} shows the output of the ReLU-max pooling network where the inputs are chosen using the Monte Carlo simulation. This output closely resembles the ReLU function, demonstrating that our network performs both max pooling and ReLU functions simultaneously. The $3\times 3$ ReLU-max pooling network consumes an average power of 17.86$\mu W$.

\section{Segmentation results}
\label{sect:segmen_results}

We evaluate our UNet design using the Cambridge-driving labeled video (CamVid) Database \cite{brostow2009semantic}. This data was captured from the perspective of a driving car, the driving scene increases the number and diversity of the observed object classes. The dataset contains 701 colored images with dimensions of $512\times 512$ pixels, each pixel is labeled into one of 32 possible classes. These classes include objects such as buildings, cars, roads, children, bicyclists, etc. To evaluate our network, we partitioned the 701 images into sets of 369 for training, 100 for validation, and 232 for testing purposes.

\begin{figure}[!ht]
	\centering
 
	\subfigure[]{\includegraphics[width=0.49\linewidth]{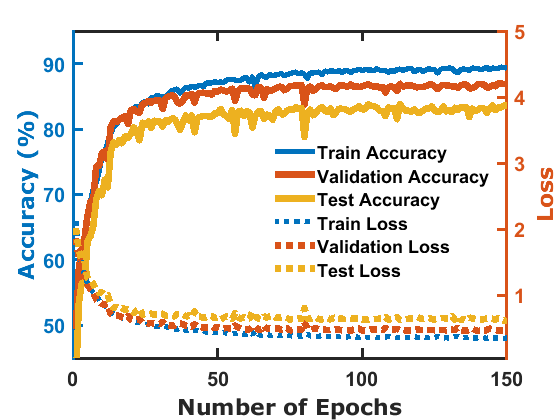}\label{fig:unet_accu}}
	\subfigure[]{\includegraphics[width=0.49\linewidth]{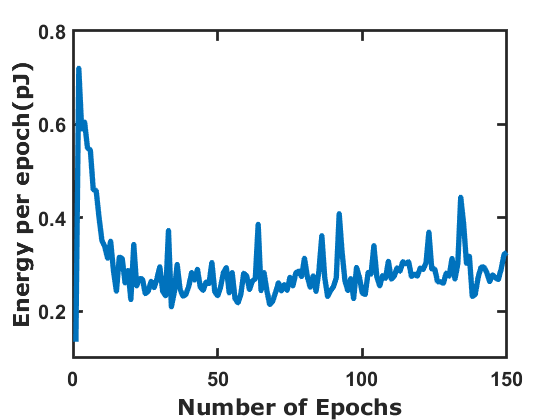}\label{fig:unet_energy}}
	\caption{(a) Accuracy($\%$) and loss of the UNet across 150 epochs for the CamVid dataset. (b) Energy consumption in all DW-synapses during network training as a function of the number of epochs.}
\end{figure}
\begin{table*}
\caption{Performance metrics for on-chip training for different thermal stability factors of ReLU and ReLU-Max pooling networks}
\begin{ruledtabular}
\begin{center}
          \begin{tabular}{m{2.5cm}m{2.5cm}m{2.5cm}m{2.5cm}m{2.5cm}m{3cm}}
          \textbf{Thermal Stability factor $\Delta$} &  \textbf{Error ($\%$) of ReLU circuit} &\textbf{Error ($\%$) of ReLU-Max pooling circuit} & \textbf{Validation accuracy for on-chip learning ($\%$)} & \textbf{Testing accuracy for on-chip learning ($\%$)} & \textbf{Total energy consumed for on-chip learning (mJ)} \\ \hline \\
        4.58 & 2.68 & 2.18  & 86.87 & 83.71 & 85.79 \\
 
        30.26 & 0.66 & 0.48 & 87.19 & 83.64 & 462.72 \\
 
         45.81 & 0.40 & 0.42 & 86.86 & 83.45 & 821.39 \\
         \end{tabular}
\end{center}
\end{ruledtabular}
\label{table:comparision}
\end{table*}
We show in Table. \ref{table:comparision} of energy consumed by the network for different thermal stability factors of ReLU, ReLU-max pooling circuits. This demonstrates a significant reduction in network energy consumption by employing ferromagnets with lower $\Delta$ values, all while maintaining segmentation accuracy. Specifically, there is a $9.57\times$ improvement in energy when utilizing a ferromagnet with a $\Delta$ of 4.58 compared to one with a $\Delta$ of 45.81.

\begin{figure}[!ht]
	\centering
    \includegraphics[width=0.60\linewidth]{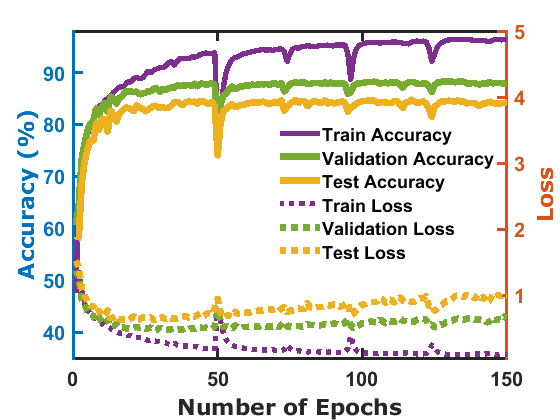}
	\caption{Software implementation of UNet, accuracy($\%$) and loss over 150 epochs for the CamVid dataset. } 
 \label{fig:unet_soft_accu}
\end{figure}
Figure \ref{fig:unet_accu} shows the accuracy($\%$) and loss of UNet over 150 epochs for testing and validation datasets. We achieved a validation accuracy of 86.87$\%$ and testing accuracy of 83.71$\%$ using ReLU, ReLU-max pooling circuits with $\Delta=4.58$, these results closely resemble those of the fully software-based implementation as shown in Fig.\ref{fig:unet_soft_accu}, where the validation accuracy is 87.95$\%$ and the testing accuracy is 84.53$\%$. 
We observe a settling time of 4ns for the ReLU emulation, and the worst-case settling time for the 9-input ReLU-max pooling network is 12ns. Considering the data path of the UNet architecture, and assuming the timings are primarily influenced by the ReLU and ReLU-max pooling networks, we estimate that the minimum time required for the input image to traverse the network to be 48ns.
We also calculated the energy consumed by the synapses during training as shown in Fig. \ref{fig:unet_energy}. The energy dissipation per epoch decreases as the network undergoes training and the weights converge. 
The total energy consumed by the network during training over 150 epochs is 85.79$mJ$, out of which 44.30$pJ$ is consumed by the synapses for weight updates.
The energy consumed by the network to process one image during testing is 1.55$\mu J$ with $\Delta$ of 4.58, this energy is dissipated in ReLU and ReLU-Max pooling units.

\begin{figure}[!ht]
	\centering
 \includegraphics[width=0.99\linewidth]{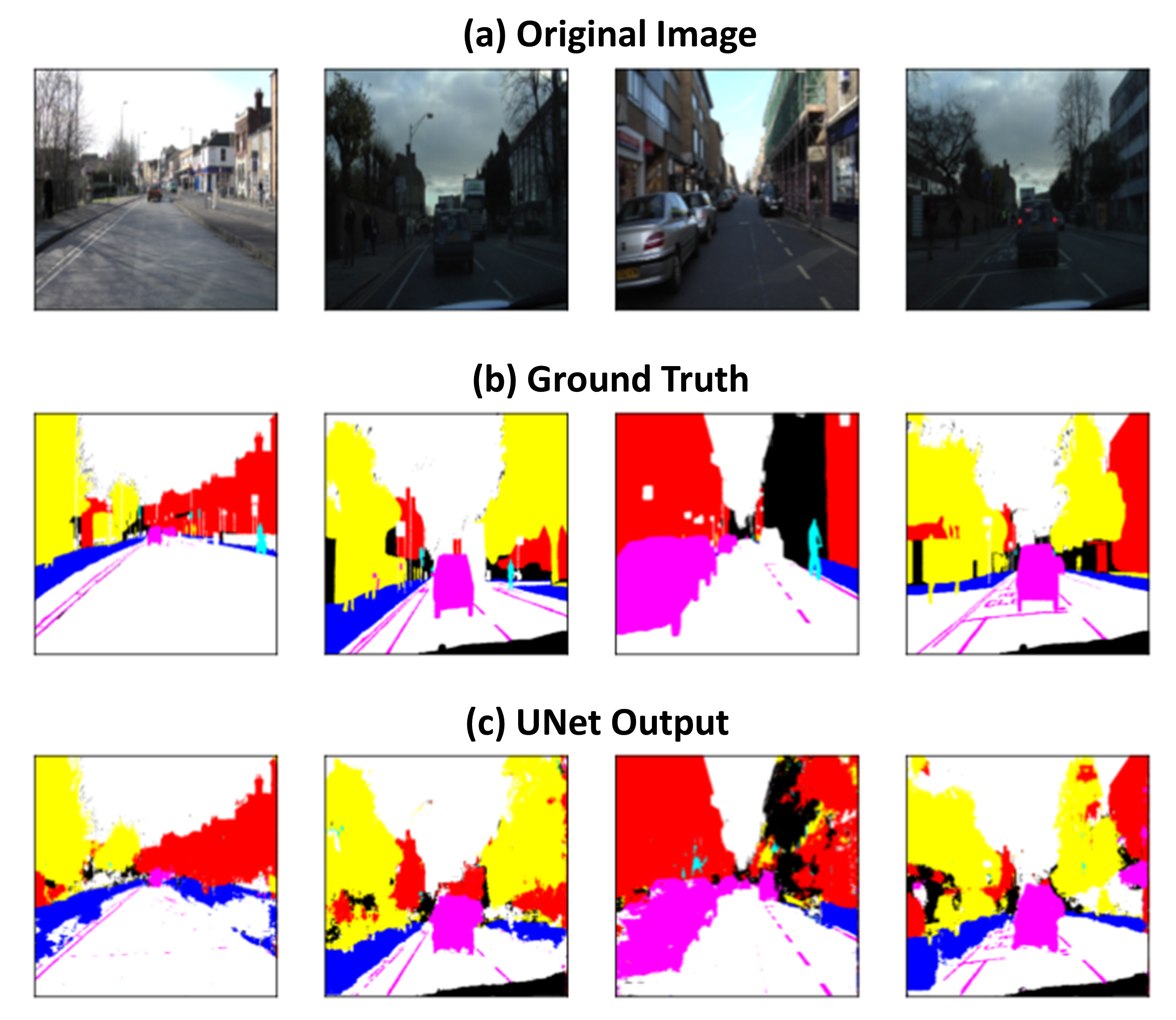}
	\caption{Comparison of UNet results to the ground truth. (a) Four images from the test set of the CamVid database. (b) Ground truth corresponding to these images. (c) Results obtained from our UNet for the four test images. Each color in the label images corresponds to a distinct element in the image; for instance, cars are represented by pink, while buildings are depicted in red.}
 \label{fig:unet_out_compa}
\end{figure}

We show in Fig. \ref{fig:unet_out_compa} the UNet output of four test images along with the ground truth labels. Here the predicted segmentation results based on our spintronic hardware implementation of UNet closely resemble the ground truth labels.

\section{Discussion}
\label{sect:discussion}
The physical realization of the UNet architecture poses challenges due to its large number of parameters.
But, to address complex problems, we need a large number of parameters, and this number will continue to grow with increasing problem complexity. Therefore, realizing these networks on specialized neuromorphic hardware is essential for efficient, scalable solutions.
Recently, significant progress has been made in the commercial realization of a very high number of spintronic devices. Some notable works include projects from Renesas, Avalanche Technology, NUMEM $\&$ IC’Alps, Everspin Technologies that have developed upto 8Gb memrories based on STT-MRAM.
Most of these are for STT-MTJ based memory realizations, but they can be extended to neuromorphic computing due to the similarity between memory architectures and cross-bar arrays \cite{xu2018stt}.

These developments in fabricating extremely large number of devices offer promising prospects for spintronics-based neuromorphic computing, yet there remains a significant journey ahead. Currently, the implementation of SHE-MTJs and domain-wall MTJs is confined to laboratory settings, requiring further time and effort to enable the integration of a large number of these devices on a chip, which would enable the implementation of complex machine learning algorithms.

\section{Conclusion}
\label{sect:conclusion}
In this article, we proposed spintronic-based hardware implementation for highly complex image segmentation tasks. 
We showcased the convolution and deconvolution designs based on domain wall MTJ and also presented the ReLU, and max pooling implementations using orthogonal current injected MTJs.
We presented our simulation platform that couples the micromagnetic simulation, NEGF, circuit simulation, and network implementation to capture the diverse physics of spin-transport, magnetization dynamics, and CMOS elements.
We demonstrated the potential of our hardware implementation of UNet by assessing its performance on the CamVid dataset, our results closely match those obtained from software implementation. We showed that employing an unstable ferromagnet for designing ReLU and max pooling functions leads to a nearly  $10\times$ reduction in network energy consumption for on-chip training, down to $85.79mJ$, without compromising segmentation accuracy.

\section*{Acknowledgments}
The author BM acknowledges the support by the Science and Engineering Research Board (SERB), Government of India, Grant No. MTR/2021/000388, and the Ministry of Human Resource Development (MHRD), Government of India, Grant No. STARS/APR2019/NS/226/FS under the STARS scheme. The author AS acknowledges the support by SERB, Grant No. SRG/2023/001327.

\subsection*{Conflict of Interest}
The authors have no conflicts to disclose.


\subsection*{Data Availability}
The data that support the findings of this study are available within the article and the CamVid dataset used in the paper is available at https://mi.eng.cam.ac.uk/research/projects/VideoRec/CamVid/. The ML codes used in this work are publicly available at https://github.com/vaddevenkatesh19/UNetPaperCodes.

\bibliographystyle{unsrt}

\bibliography{reference}

\end{document}